\documentclass{article}
\usepackage{kmn}
\usepackage{epsfig,subfigure}
\title[The Galactic Disk Distribution of Planetary Nebulae with Warm
Dust Emission Features: II]{The Galactic Disk Distribution of
Planetary Nebulae with Warm Dust Emission Features: II}

\author[S.~Casassus \& P.\,F.~Roche]
  {S.~Casassus$^{1,2}$, P.\,F.~Roche$^1$ \\
	$^1$ Astrophysics, Physics Department, Oxford University, Keble Road, Oxford OX1 3RH\\
	$^2$ Departamento de Astronom\'{\i}a, Universidad de Chile, Casilla 36-D, Santiago, Chile.}
\date{Accepted ... Received ...}
\pagerange{\pageref{firstpage}--\pageref{lastpage}}
\pubyear{2000}
\def\gs{\mathrel{\raise1.16pt\hbox{$>$}\kern-7.0pt
\lower3.06pt\hbox{{$\scriptstyle \sim$}}}}
\def\ls{\mathrel{\raise1.16pt\hbox{$<$}\kern-7.0pt
\lower3.06pt\hbox{{$\scriptstyle \sim$}}}}
\begin{document}
\label{firstpage}
\maketitle
\begin{abstract}
We address the question of whether the distribution of warm-dust
compositions in IR-bright galactic disk PNe (paper I, Casassus et
al. 2000), can be linked to the underlying stellar population. The PNe
with warm dust emission represent a homogeneous population, which is
presumably young and minimally affected by a possible dependence of PN
lifetime on progenitor mass. The sample in paper~I thus allows testing
the predictions of single star evolution, through a comparison with
synthetic distributions and under the assumption that tip-of-the-AGB
and PN statistics are similar.  We construct a schematic model for AGB
evolution (adapted from Groenewegen \& de Jong 1993), whose
free-parameters are calibrated with the luminosity function (LF) of C
stars in the LMC, the initial-final mass relation, and the range of PN
compositions. The observed metallicity gradient and distribution of
star forming regions with galactocentric radius (Bronfman et al. 2000)
allow us to synthesise the galactic disk PN progenitor population. We
find the fraction of O-rich PNe, f(O), is a tight constraint on AGB
parameters.  For our best model, a minimum PN progenitor mass
$M^\mathrm{min}=1$~M$_{\odot}$ predicts that about 50\% of all young
PNe should be O-rich, compared to an observed fraction of 22\%; thus
$M^\mathrm{min}=1.2$~M$_{\odot}$, at a $2\sigma$ confidence level
($M^\mathrm{min}=1.3$~M$_{\odot}$ at $1\sigma$). By contrast, current
AGB models for single stars can account neither for the continuous
range of N enrichment (Leisy \& Dennefeld 1996), nor for the
observation that the majority of very C-rich PNe have Peimbert type~I
(paper~I). f(O) is thus an observable much easier to model.  The
decrease in f(O) with galactocentric radius, as reported in paper~I,
is a strong property of the synthetic distribution, independent of
$M^\mathrm{min}$. This trend reflects the sensitivity of the surface
temperature of AGB stars and of the core mass at the first thermal
pulse to the galactic metallicity gradient.
\end{abstract}
\begin{keywords}
planetary nebulae: general -- infrared: ISM: lines and
bands -- ISM: abundances  -- stars: evolution -- stars: AGB and post-AGB.
\end{keywords}

\section{Introduction}
\label{sec:intro_4}

The vertical stratification of planetary nebulae (PNe) according to
their N/O abundance ratio (i.e. the Peimbert 1978 classification), is
an indication that PN compositions can be linked to the progenitors'
masses. Yet few attempts have been taken at building a comprehensive
picture of the distribution of PN chemical compositions in the
galactic context, although the range of C, N and O abundances has been
successfully modelled (e.g. Kaler et al. 1978, Renzini \& Voli 1981,
Bryan et al. 1990 , Groenewegen et al. 1995, Henry et
al. 2000). The difficulties in PN population synthesis stem from an
incomplete understanding of the late stages of stellar evolution, even
for single stars on the asymptotic giant branch (AGB). Additionally,
knowledge of the observational properties of PNe suffer from the lack
of accurate distances, and the variety of object types and
ages.

In paper I (Casassus et al. 2000) we reported on the 8--13~$\mu$m dust
features in catalogued compact and IR-bright galactic disk PNe. The
PNe with warm dust emission represent a homogeneous population: their
maximum age is limited by the presence of dust close to the central
star.  The sample is thus minimally affected by a possible dependence
of PN lifetime on progenitor mass (keeping in mind that selection
effects in the direction of differential `warm-dust' lifetimes as a
function of grain composition are difficult to quantify, see Section~6
in paper~I and Section~\ref{sec:linking} in this work). The dust-grain
spectral signatures also provide a diagnostic 
of the C/O chemical balance, and an alternative to the gas-phase
abundances (e.g. Barlow 1983, Roche 1989). There is a trend for a
decreasing fraction of O-rich PNe outside the solar circle, reflecting
the decrease in the ratio of M to C stars found by Thronson et
al. (1987) and Jura et al. (1989).

Can the current scenarios for single star evolution reproduce the
observed frequencies of O- or C-rich grains in PNe, and the
frequencies of Peimbert (1978) types among grain-types? Does the
metallicity dependence of AGB evolution models predict the trends with
galactocentric radius? In this article, we analyse the galactic
distribution of PNe compositions in terms of simple stellar evolution
models.  We employ a set of analytical prescriptions in order to
evolve stars with a continuous range of initial masses
($M_\mathrm{i}$, from 1 to 7~M$_{\odot}$) and initial metallicity
($Z_\mathrm{i}$, from 0.005 to 0.035).  This synthetic AGB model,
coupled with a simple model for the galactic disk, allows us to check
whether single star evolution can reproduce the distribution of PN
chemical abundances.

We will conclude that the fractions of PNe displaying the dust
signatures of O-rich and C-rich grains,
$\mathrm{f(O)}+\mathrm{f(C)}=1$, are tight constraints on the
parameters of AGB evolution models. Synthetic AGB models that account
for f(O) also reproduce the C star luminosity function (LF) in the
LMC; but the reverse is not true. However, for the PN sample analysed
here, the minimum progenitor mass needs to be adjusted -- our analysis
favours a minimum progenitor mass of 1.2~M$_\odot$ at 2$\sigma$. The
galactic f(O) gradient stems from the metallicity gradient, through
the prescriptions for the core mass at the first thermal pulse and the
AGB surface temperature. These results are an incentive to perfect the
modelling, and pursue the use of PNe as probes of their progenitor
population in different galactic environments. Also, the advantage of
PNe for testing AGB evolution models resides in their well defined
evolutionary status. For comparison, it is difficult to ascertain the
TP-AGB membership of M stars, making the trends in the C/M star ratio
much more difficult to model (see Section~\ref{sec:lmctest}).

In Section \ref{sec:sagb} we present the procedure followed to
synthesise the galactic PN population, together with a brief
description of the model used to evolve stars along the AGB (complete
details are given in the Appendix).  The free parameters in the AGB
model were calibrated with observed quantities; this synthetic AGB
model should thus be taken as an observational requirement rather than
a detailed calculation. We investigate whether a model that
reproduces the C star LF in the LMC also reproduces f(O). Section
\ref{sec:linking} describes the process of linking the model results
with the PN sample, the results of which will be presented in
Sect. \ref{sec:min_mass}, together with an estimate of the minimum
mass required to model the population of PNe with warm dust emission
features.  Sect. \ref{sec:conc_4} contains a summary of our
conclusions.

\section{Synthetic distribution of PNe compositions in the galactic disk}\label{sec:sagb}

\subsection{Synthetic AGB model}
\label{sec:sagb_params}

We aim at testing existing AGB evolution scenarios, such as that
proposed by Groenewegen \& de Jong (1993, GJ93). Synthetic AGB
evolution models use empirical laws derived from detailed stellar
evolution codes.  As a starting point, we used GJ93, who follow
Renzini and Voli (1981, RV81) and include an explicit metallicity
dependence whenever possible.  Other models exist, like those of Bryan
et al.  (1990) and Marigo et al.  (1996, 1999), but they have not been
tested as extensively against observations as that of GJ93, which is
also very light on computer resources and readily reproducible. The
model is primarily concerned with the thermally pulsing AGB (TP-AGB),
while the initial conditions at the start of the TP-AGB are functions of
$(M_i,Z_i)$.

The set of analytical prescriptions was taken from the compilation by
GJ93, although we also considered a different treatment of hot bottom
burning (HBB), making use of the results in Forestini \& Charbonel
(1997, FC97)\footnote{In GJ93, the density at the bottom of the
convective envelope was taken as a constant 2~g~cm$^{-3}$ (Martin
Groenewegen, private communication), while these values are reached
only in the most massive models of FC97}.  The free parameters which
bear directly on f(O) are $\eta_\mathrm{AGB}$, the mass loss parameter
on the TP-AGB for a scaled Reimers (1975) law, and $\lambda$, the
fraction of core growth between two consecutive thermal pulse that is
dredged-up and mixed.  We kept fixed the other parameters in GJ93: the
minimum core mass for 3$^\mathrm{rd}$ dredge-up,
$M_{c}^\mathrm{min}=0.58\,$M$_{\odot}$; the mass loss parameter for the
red giant branch, $\eta_\mathrm{RGB}=0.86$; the mass loss parameter
for the early-AGB, $\eta_\mathrm{EAGB}=3$.

The model was required to reproduce the C star LF in the LMC and the
initial-final mass relation for stars in the solar neighbourhood, as
in GJ93 (Sect.~\ref{sec:app_obs}). The best model, which fits the C
star LFs from Costa \& Frogel (1996) and GJ93, corresponds to
$\lambda=0.75$, $\eta_\mathrm{AGB}=4$, with a mass-loss rate high
enough to keep the final masses within the observed limits.  However,
while an equally good fit to the GJ93 C star LF is obtained with
$\lambda=0.6$, $\eta_\mathrm{AGB}=5$, this is rejected by f(O)
(Section~\ref{sec:model_sensitivity}).

\begin{figure}
\begin{center}
\resizebox{8cm}{!}{\epsfig{file=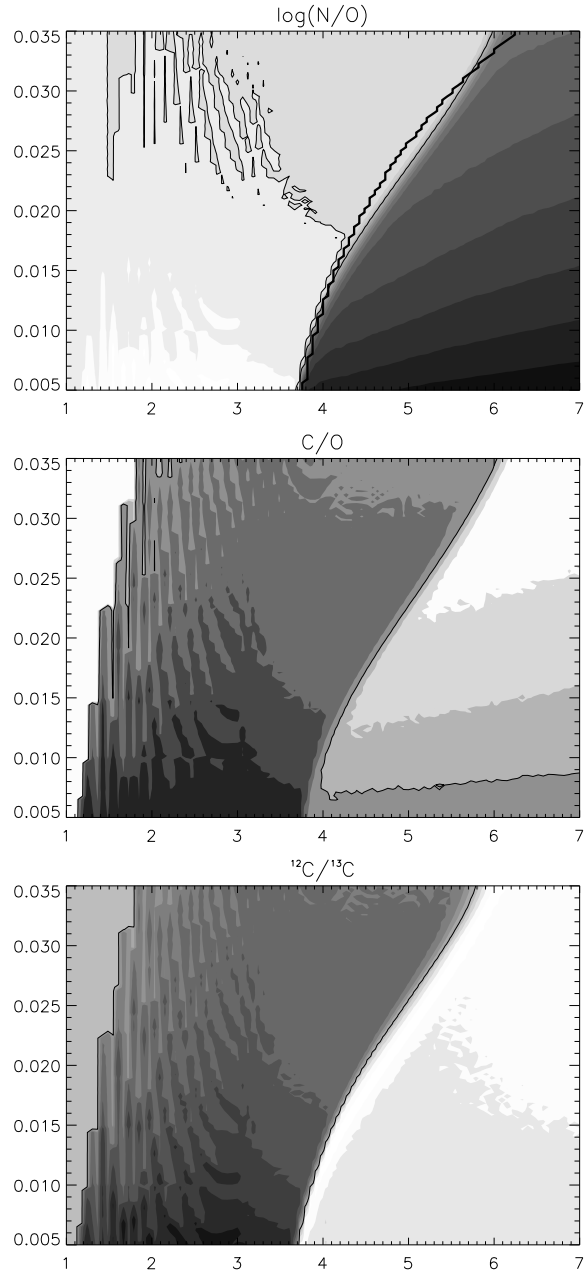}}
\end{center}
\caption{ PN abundances averaged over the last 25\,000~yr of AGB
evolution as a function of initial mass in x, initial metallicity in
y. From top to bottom: {\bf log(N/O)}, with darker grey scale contours
at --0.8, --0.7, ..., 0.6, and the thin solid lines at --0.6 and
--0.3 represent the contours for type IIa and type I PNe, while the
thick line traces the minimum initial mass for the occurrence of the
second dredge-up (a sharp transition at $-0.3$ results in missing
contour levels, but the minimum and maximum plotted values of the grey
scale are --0.8 and 0.6); {\bf C/O}, with darker contours at
$\log($C/O$)$= --0.6, ..., 0.4, 0.6, and the solid line is at C/O=1;
{\bf $^{12}$C/$^{13}$C}, darker contours at $\log$($^{12}$C/$^{13}$C)=
0.7, 0.9,  ..., 2.9, and the solid line is at
$^{12}$C/$^{13}$C=30.}\label{fig:mz}
\end{figure}

The resulting PN abundances ($\lambda=0.75$, $\eta_\mathrm{AGB}=4$),
as a function of initial mass and initial metallicity are shown in
Figure~\ref{fig:mz}. The set of abundances summarised in
Fig.~\ref{fig:mz} will be referred to as our standard model. The
stellar ejecta were averaged over the last 25000~yr, to compare with
gas-phase abundances. The different PN classes based on the relative
abundances of $^{12}$C, $^{13}$C, N, O have their niche: O rich PNe
are expected for both the lowest and highest $M_i$ (type I PNe
occurring for the highest progenitor masses), and $^{12}$C/$^{13}$C
ranges from CN equilibrium values ($\sim$3) to $>$100.

If the stellar ejecta is averaged over the last 2000~yr only, which is
taken to represent warm-dust compositions (see
Section~\ref{sec:linking}), the contours for PN compositions are still
the same as in Fig.~\ref{fig:mz} in the case of $\log$(N/O) (albeit a
noisier IIa contour). In the cases of C/O and $^{12}$C/$^{13}$C the PN
composition map as a function of $(M_i,Z_i)$ is also similar to
Fig.~\ref{fig:mz}, but the point to point variations are sharper and
reach higher peak value ($\sim$13 and 3000 at $t_\mathrm{PN}=2000$~yr
against $\sim$7 and 900 at $t_\mathrm{PN}=25000$~yr, for C/O and
$^{12}$C/$^{13}$C, respectively). There is one important change
between $t_\mathrm{PN}=2000$~yr and $t_\mathrm{PN}=25000$~yr, however:
in terms of the PN compositions, for $t_\mathrm{PN}=2000$~yr the
effects of HBB for masses higher than 4--5~M$_{\odot}$ are not as
dramatic, and C/O$>1$ even for the most massive progenitors (see Frost
et al. 1998 for a physical discussion on the quenching of HBB).

\subsection{Problems with N enrichment in PNe} \label{sec:pb_N}

In order to reproduce the range of N and C enrichment found in PNe and
to account for the observed frequencies of type I and IIa PNe, we find
that HBB must be much more efficient than currently modelled.  As
noted by Leisy \& Dennefeld (1996), there is continuity in the
nitrogen abundances of PNe in the LMC, with no particular grouping
above the type I threshold. And yet at Z=0.008, typical of the LMC,
the synthetic AGB model predicts no type IIa PNe at all (in the
Fa\'{u}ndez-Abans \& Maciel 1987 classification,
i.e. $-0.6<\log(\mathrm{N}/\mathrm{O})<-0.3$). These conclusions also
hold for the GJ93 model: with their treatment of HBB, Fig.~\ref{fig:mz}
is largely unchanged (except perhaps at $Z_i \sim 0.03$, where the
type~I threshold is at 4.6~M$_{\odot}$ instead of
5.5~M$_{\odot}$). Models in which core-growth on the AGB is
negligible, and where HBB effectively occurs above a critical core
mass, predict a sharp transition in N/O ratio and progenitor mass,
which is not observed.

Additionally, the majority of very C-rich PNe (those whose mid-IR
spectra display the family of emission bands associated with PAHs) are
of type I (paper~I). However,  the existing synthetic AGB models cannot
produce N-rich and very C-rich PNe in the observed frequency.
Fig.~\ref{fig:mz} shows how the highest C enrichment should be found
for PNe with solar N/O ratios (again, the PN composition map obtained
with a 2000~yr average of AGB mass-loss leaves Fig.~\ref{fig:mz}
largely unchanged). In terms of the gas-phase abundances, type I PNe
are divided in their C/O chemical balance (Leisy \& Dennefeld 1996,
their Fig.~12; see also Aller \& Czyzak 1983 for galactic disk
PNe). But the sharp onset of HBB with $M_i$ produces O-rich PNe,
exclusively (for $Z_i \gs 0.01$).

An exception to the above problems with synthetic AGB models is
provided by RV81.  It appears that RV81 underestimated the mass loss
rate ($\eta_\mathrm{AGB}\sim1/3$ predicts excessive final masses, GJ93
and references therein), and the extended lifetime allows the
temperature at the bottom of the convective envelope to rise
significantly with the growth of the core during the TP-AGB. A
$3.3$~M$_\odot$ star can start dredging-up carbon and then undergo HBB
later in its evolution, in a balance that produces C and N rich
PNe. Thus the problems with C and N enrichment in PNe could be solved
by artificially fixing the slope for the increase of HBB temperature
with core-growth, $dT_\mathrm{B}/dM_c$, to the values found by FC97
for their 5--6\,M$_{\odot}$ models. We found that models with
$dT_\mathrm{B}/dM_c=2\,10^{9}$\,K\,M$_\odot^{-1}$ produce type~I and
IIa PNe in the observed frequencies (around 30\% and 20\% in the
sample of paper~I, see also Maciel \& Dutra 1992), with type~I PNe
being preferentially very C-rich. An interesting alternative to a high
$dT_B/dM_c$ slope can be found in Siess \& Livio (1999). The accretion
of brown dwarfs or giant planets could trigger CNO processing in the
convective envelope.  Although the concept of envelope heating from
the kinetic energy of accreted material is appealing, it is not
obviously related to the observed N/O latitude effect.

Thus the N/O classification of PNe, perhaps one of their most widely
used chemical characteristic, turned out to be impossible to use in PN
population synthesis, at least with the models we tried. On the other
hand, f(O) depends mostly on $M_{c}^\mathrm{min}$ coupled with the
core mass at the first thermal pulse (in this formalism). In fact,
above a threshold initial mass, all PNe are predicted to generate
C-rich warm dust during the earliest stages of their evolution.
f(O) is therefore a much easier quantity to model.



\subsection{Synthetic distribution of PN progenitors in the galactic disk}\label{sec:diskmodel}

A set of 10000 stars was evolved to construct the synthetic
distribution of PNe in the galactic disk. The probability distribution
for the star formation rate as a function of galactocentric radius was
extrapolated from the ultra-compact H\,{\sc ii} region survey in
Bronfman et al. (2000). We assumed the number density of massive star
forming regions would also trace the number density of newly born PN
progenitors.  Initial stellar masses were distributed according to an
initial mass function (IMF) index of $-1.72$ (in $dN/d\log M$). This
index is also used in the derivation of the LMC C star LF
(Section~\ref{sec:lmctest}), and is the same as in GJ93.  The lower
mass cutoff for the compact PNe progenitors was kept as a free
parameter, $M^{\mathrm{min}}$, and the upper mass limit was
7\,M$_{\odot}$.  The IMF index may seem steep for intermediate mass
stars, and we investigated using $-0.95$, as derived in Sabas (1997)
from a complete sample of B5-F5 stars and hipparcos parallaxes (over
1.2--4 M$_{\odot}$). In this case $M^{\mathrm{min}}$ turned out to be
somewhat lower (see Section~\ref{sec:min_mass}).

Radial diffusion in the galactic disk was taken into account following
the results presented in K\"{o}ppen and Cuisinier (1997), in the
epicyclic approximation.  The age-velocity relation by Wielen (1977),
\begin{equation}
\sigma_v=\sqrt{100+600\frac{\mathrm{age}(M)}{10^9\,yr}}\,\mathrm{km\,s^{-1}},
\end{equation}
in conjunction with a velocity ellipsoid of constant shape (Wielen
1977),
\begin{equation}
\sigma_\mathrm{U}:\sigma_\mathrm{V}:\sigma_\mathrm{W}= 0.79 : 0.46 :0.41,
\end{equation}
allows to estimate the spread of the normal distribution by which
final galactocentric radii are distributed (Wielen et al. 1996),
\begin{equation}
\sigma_\mathrm{R}(\mathrm{age})=1.56\,\sigma_\mathrm{U}(\mathrm{age})/\kappa,
\end{equation}
where $\kappa=31.6$\,km\,s$^{-1}$\,kpc$^{-1}$ is the epicyclic
frequency of stellar orbits in the disk, at $R=R_\circ$.  The factor
1.56 could altogether be neglected in this schematic treatment, which
will be applied to $\approx .5-1.5\,R_{\circ}$.

As argued in Carraro et al. (1997), we adopted a radial metallicity
gradient of $-0.07$~dex/kpc and constant in time, and assumed a 0.005
FWHM scatter when drawing initial metallicities.  $Z(R=R_{\circ},t)$
was required to match the age metallicity relation by Meusinger et
al. (1991),
\begin{eqnarray}
[\mathrm{Fe}/\mathrm{H}](t)& \approx & \log\left[
Z(R=R_{\circ},t)/Z_{\odot} \right] = \\
\nonumber & & \log \left[1.98-1.88(1-t/28.5)^{1.25}\right].
\end{eqnarray}

\section{Linking the modelled and observed PN galactic distributions}\label{sec:linking}

We based the comparison between modelled and observed distributions on
the sample of compact and IR bright PNe from paper~I. In linking the
modelled and observed PN galactic distributions, the selection
effects in the PN catalogues need to be taken into account, as well
as the exact relation between the AGB stellar ejecta and the observed
PN compositions.

The PNe in the sample with warm dust emission were required to have {\it
IRAS\/} 12\,$\mu$m band fluxes in excess of 0.5\,Jy. Their
distribution was modelled by applying the same selection criterion to
the synthetic PN population. It is shown in paper~I that the PNe with
warm dust should be optically thick in the Lyman continuum, in which
case the four {\it IRAS\/} bands give a good representation of their
bolometric flux, with a fraction of $0.25\pm0.14$ of the total flux in
the 12$\mu$m band. The IR-bright selection criterion should probe the
whole of the galactic disk. But selection effects that limit the
galactic disk sampling area come from the optical discovery
surveys. Extinction in the galactic disk needs to be taken into
account. Condon \& Kaplan (1999) report extinction values towards 429
PNe; no PNe are found with
$c=\log(\mathrm{H}\beta^\mathrm{real}/\mathrm{H}\beta^\mathrm{obs})$
in excess of 3.3, with an approximate cutoff in the distribution at
$c\sim2$. They also show that $c$ depends strongly on galactic
coordinates, with values of order 1 towards the galactic
anticentre. The bulk of the extinction to PNe is thus galactic, rather
than intrinsic, and a maximum value of c=2--3.5 would give
$A_\mathrm{V}=3.8-6.6$. We adopted a maximum interstellar extinction
of $A_\mathrm{V}=4$, as for the objects with warm dust intrinsic
extinctions of at least one magnitude have been reported.

A gaussian profile\footnote{K\"{o}ppen \& Vergely (1998) modelled the
galactic disk extinction by matching an exponential linear density of
extinction in R and z to bulge PNe. But the dust and molecular content
at $R\ls2$ is similar to that of the molecular ring at
$R\sim0.5\,R_{\odot}$ (Bronfman et al. 1988, Deul \& Burton 1991), and
it is not clear whether there is a sharp central concentration of dust
in the galactic disk.}  in galactocentric radius for $\Lambda(R)$, the
`linear density of extinction', gives
\begin{equation}
A_\mathrm{V}(D,l) = \int_0^{D}\,ds\,\Lambda(R) =
\int_0^{D}\,ds\,A\,\exp(-\frac{R^2}{2\,R_{\mathrm{A}}^2}).
\end{equation}
The values of $A=4$~mag\,kpc$^{-1}$, $R_{A}=3.6\,$kpc were fixed by
requiring $A_\mathrm{GC}=20$~mag for the extinction to the galactic
centre, and a minimum galactocentric radius of 0.5\,$R_{\circ}$ for
$A_\mathrm{V}^\mathrm{max}$=4. These requirements give a local linear
density of visual extinction of only 0.25~mag\,kpc$^{-1}$, and the
value of $A_\mathrm{GC}=20$~mag may seem low as well. However, we did
not take into account the dependence of $\Lambda(R)$ on height above
the galactic plane, and that PNe would preferentially be discovered
towards directions of low extinction: if the average linear density of
extinction to local PNe were really 1~mag\,kpc$^{-1}$, no PNe would be
detected beyond 4\,kpc.  It is apparent from Fig.~6 in paper~I that
the maximum distance to galactic disk PNe with warm dust is about
8\,kpc, which we adopted as a maximum distance in the synthetic
population.  This estimate of the sampled area in the galactic disk is
intented to approximate the observed face-on distribution.


In the case of the PNe with warm dust emission features, the problem
of a lifetime dependence on the progenitor mass is greatly simplified
by the presumed youth of the objects. We assumed that the compact and
IR-bright PNe with warm dust emission all have the same lifetime,
neglecting any dependence on $M_\mathrm{i}$. But it should be kept in
mind that the durations of the `warm-dust' phase lifetimes could
depend on the dust opacities. It was suggested to us (Kevin Volk,
private communication) that this could be the reason why the fraction
of O-rich objects is f(O)$\sim$22\% from the dust grain compositions,
while it is closer to 40\% from the plasma diagnostics.  Some
arguments against this possibility may be found in Section~6 of
paper~I, and the next section develops the idea that the difference in
f(O) could be due to the different timescale over which the AGB ejecta
should be averaged ($t_\mathrm{PN}$ in this work). In any case, the
conclusions we reach on f(O) being a tight constraint on AGB
parameters are valid even if f(O)$\sim$50\%.

A lower metallicity limit for the production of dust is discarded, as
even some very low metallicity PNe show warm-dust emission (e.g. in a
forthcoming article on bulge PNe, we will report on the type~IV PN
M~2-29, that shows the silicate signature in spite of a very low O
abundance of [O/H]$=-1.4$~dex, as estimated by Ratag et al. 1997).


Since PNe with no warm dust correspond to later evolutionary stages
(paper~I), the link between the computed C/O ratios and the dust
features is very tight. As discussed in paper~I, Silicates represent
C/O$<$1, and SiC and UIR bands C/O$>$1. The only problem left in
linking the synthetic and observed PN populations is defining how to
compute the average compositions. If the warm dust grains are heated
directly by radiation from the central stars, then a temperature of
100--200\,K places the grains at $\sim10^{16}$\,cm from the exciting
star, which corresponds to an ejection age of 400\,yr at
10\,km\,s$^{-1}$. In this picture, the warm dust grain composition
derives from the very last stages of AGB evolution.  Thus the PNe
compositions were averaged over the last 2000~yr, to compare with the
dust emission features, while a more topical 25\,000~yr was used to
compare with the gas phase abundances.

\begin{figure}
\begin{center}
\resizebox{8cm}{!}{\epsfig{file=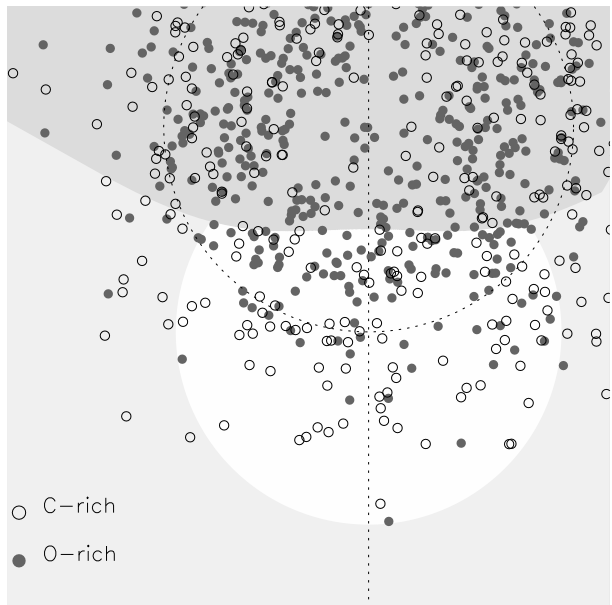}}
\end{center}
\caption{Synthetic distribution of PNe in the galactic disk, with
filled circles representing O-rich nebulae, while open circles
represent C-rich nebulae. The dotted lines correspond to the solar
circle (R$_{\circ}$=8.5~kpc) and to $l=0$.  The clear area is the region
where sources would have been included in the PN catalogues, within
the adopted limit of $A_\mathrm{V}=4$ and $d_\mathrm{max}=8$~kpc. The
AGB mass loss was averaged over the last 2000\,yr of evolution. For
the clarity of the plot, the initial number of PNe is only 1000, and
comes down to $\sim$200 within the galactic disk sampling area. This
Figure can be compared to Fig.~6 in paper~I.}\label{fig:synt_distr}
\end{figure}

The resulting synthetic distribution can be seen in Figure
\ref{fig:synt_distr}, in the case of a minimum progenitor mass of
1.2\,M${_\odot}$. Galactocentric azimuths were drawn from a uniform
random distribution.

\section{Results: The minimum mass for PN progenitors}\label{sec:min_mass}

\subsection{The minimum mass for PN progenitors}

Having calibrated all the free-parameters in the synthetic AGB model,
the mass range for PN progenitors still needs to be specified. The
lower mass limit has a direct influence on the synthesised PN
distribution, although the exact upper mass limit is of secondary
importance, high mass stars being less frequent.

\begin{figure}
\begin{center}
\resizebox{8cm}{!}{\epsfig{file=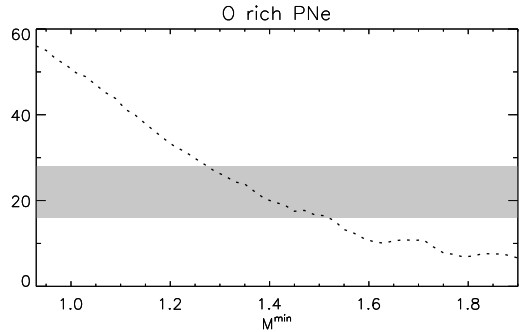}}
\caption{The fraction of O-rich PNe, f(O), as a function of minimum
main sequence progenitor mass, $M^\mathrm{min}$.  The shaded area
correspond to the observed values within $1\sigma$.}
\label{fig:results}
\end{center}
\end{figure}

The property of the synthetic distribution that can be most directly
compared to the observed distribution of PNe with warm dust is the
relative frequency of C-rich and O-rich nebulae.  Shown in
Figure~\ref{fig:results} is the fraction of O rich PNe as a function
of minimum progenitor mass, $M^{\mathrm{min}}$, for the set of
parameters given in Section~\ref{sec:sagb_params}.  PN compositions
were computed by averaging the stellar mass loss over the last
$t_{\mathrm{PN}}=2000$~yr of evolution on the AGB to compare with the
dust signatures. The shaded areas correspond to the range of observed
values within $\pm\,1\,\sigma$ (from paper I),
\begin{equation}
\mathrm{f(O\,rich)}  =  0.22\pm0.06. 
\end{equation}
In order to match the observations, $M^\mathrm{min}$ is
constrained to
\begin{equation}
M^\mathrm{min} > 1.2\,\mathrm{M}_{\odot}
\end{equation}
at two-$\sigma$ confidence level (or $\sim$1.3\,M$_{\odot}$ at
one-$\sigma$). The IMF slope in our standard model is rather steep,
and for comparison a shallow IMF of index $-0.95$ (Sabas 1997), gives
$M^\mathrm{min} > 1.2\,\mathrm{M}_{\odot}$ at one-$\sigma$ - f(O) does
not seem very sensitive on the IMF.

We will now test the value of $M^\mathrm{min}$ against less accurate
properties of the observed PN distribution, and analyse trends with
galactocentric radius.  In the synthetic PN distribution, the relative
number of PNe inside and outside the solar circle is
$N_\mathrm{in}/N_\mathrm{out}=1.3$ (rather independent of
$M^\mathrm{min}$), while the observed ratio is 0.9--1.6 (from Tables~6
and 7 in paper~I), depending on the PN distance scale. The approximate
agreement is an indication that the galactic disk sampling area of
Section~\ref{sec:linking} is appropriately defined.

\begin{figure}
\begin{center}
\resizebox{8cm}{!}{\epsfig{file=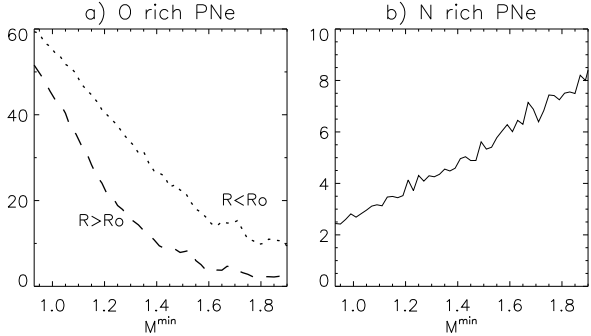}}
\caption{Properties of the synthetic PN population as a function of
minimum main sequence progenitor mass, $M^\mathrm{min}$. a) f(O)
inside and outside the solar circle, b) the fraction of type I PNe
(the observed values is $29\pm6$\%, from paper~I and references
therein).}\label{fig:results2}
\end{center}
\end{figure}

Figure~\ref{fig:results2}a shows the fraction of O rich PNe, f(O), for
$R<R_{\circ}$ and $R>R_{\circ}$. There is a clear decrease in the
model f(O), for any value of $M^\mathrm{min}$, and it is thus a robust
property of the synthetic distribution.  From paper~I, the observed
f(O) is $30\pm9$\% inside the solar circle, and $14\pm7$\%
outside. Figure~\ref{fig:results2}b shows the proportion of type I PNe
as a function of minimum progenitor mass. For any $M^\mathrm{min}$,
the fraction of type~I PNe is far too low. This result can also be
reached by estimating the proportion of progenitors between
$M^\mathrm{min}$ and $\sim 4$~M$_{\odot}$, even for a shallow
IMF. Either this is yet another deficiency of current models for N
enrichment, or the PN sample is dramatically affected by biases
against low mass progenitors. The latter case seems rather contrived;
a complete PN sample is a working hypotheses for this article, and in
the event of a strongly biased sample, $M^\mathrm{min}$ represents a
lower mass limit above which we trust the observed f(O) - as well as
f(type~I). Additionally, we know N enrichment is not satisfactorily
modelled (e.g. Section~\ref{sec:pb_N}).

For comparison the gas phase abundance distribution synthesised with
$t_{\mathrm{PN}}=25000$~yr predicts 5--12\% more O-rich PNe, depending
on $M^\mathrm{min}$, and around 28\% for
$M^\mathrm{min}=1.4$~M$_{\odot}$ (which gives f(O)=22\% if
$t_{\mathrm{PN}}=2000$). This is due to a contribution of O-rich
objects produced by HBB, which is quenched over the last 2000~yr.  In
the sample of paper~I, this fraction is $40\pm8$\%, and is $\sim40$\%
in Zuckerman and Aller (1986, note that their sample is inhomogeneous,
with a broad range of PN ages). The gas phase C/O ratio is available
for an even more restricted number of sources (of the ones with warm
dust), and are subject to large uncertainties stemming from the
difficulties in measuring the C abundance, especially in compact and
IR-bright PNe - compare for instance Aller \& Czyzak (1983) and Henry
et al. (2000), in the cases of NGC~3242 and NGC~6826. Considering such
uncertainties, we believe the difference between the fraction of
O-rich PNe as inferred from the dust compositions and the plasma
diagnostics is due to $t_{\mathrm{PN}}$.

In the framework of an artificially high $dT_\mathrm{B}/dM_c$
(Section~\ref{sec:pb_N}), the effects of $t_{\mathrm{PN}}$ on f(O) are
stronger because HBB occurs for lower progenitor masses and is thus
more frequent. Accounting for the observed proportions of type I and
IIa PNe ($29\pm6$\% and $12\pm6$\%, paper~I) would give an upper limit
to $M^\mathrm{min}$ of about 1.4~M$_\odot$ (above this limit, there
are too many type~I PNe).  With $M^{\mathrm{min}}=1.2$~M$_{\odot}$ and
$t_{\mathrm{PN}}=2000$~yr, the fraction of C-rich PNe that have
Peimbert type I is 35\%, and the observed proportion of type I PNe
among PNe with C-rich dust grains is $37\pm10$\%
(paper~I). Similarly, in the synthetic distribution the fraction of
O-rich PNe that have Peimbert type I is 26\%, and the observed
proportion of type I PNe among those with O-rich dust is $22\pm14$\%.


\subsection{Model sensitivity} \label{sec:model_sensitivity}

What is the model sensitivity of the predicted fraction of O-rich PNe?
Of the relations used in the synthetic AGB model, which are the ones
responsible for the galactic gradient in f(O), i.e. which are the
most sensitive on the galactic metallicity gradient?

\begin{figure*}
\centering
\mbox{\subfigure{\epsfig{figure=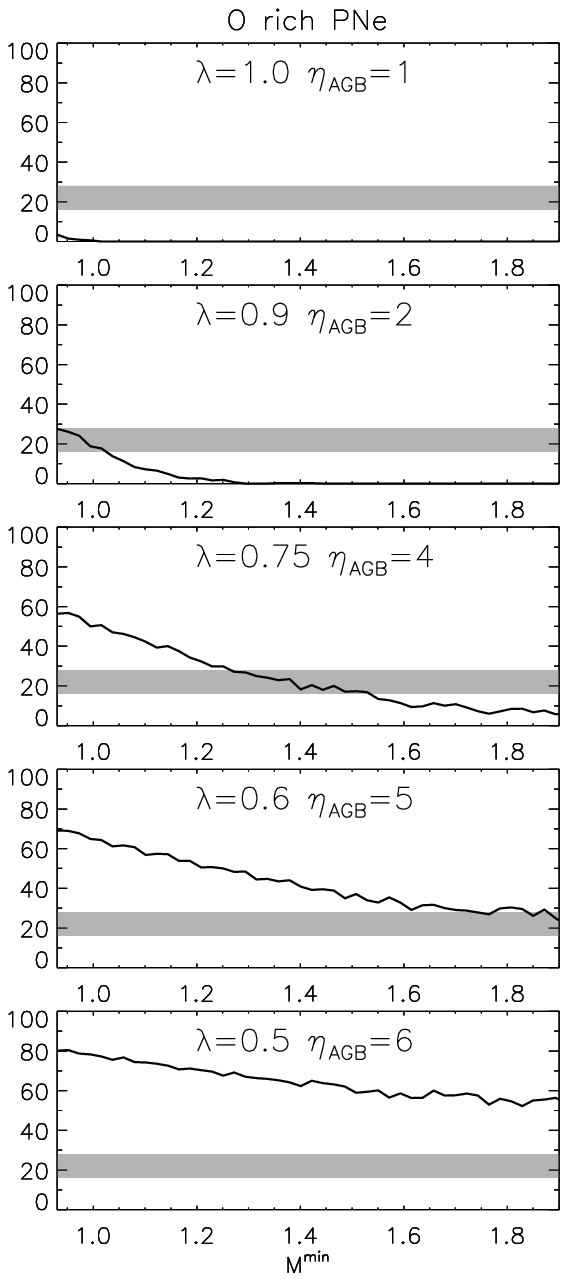,width=5.cm}}\quad
\subfigure{\epsfig{figure=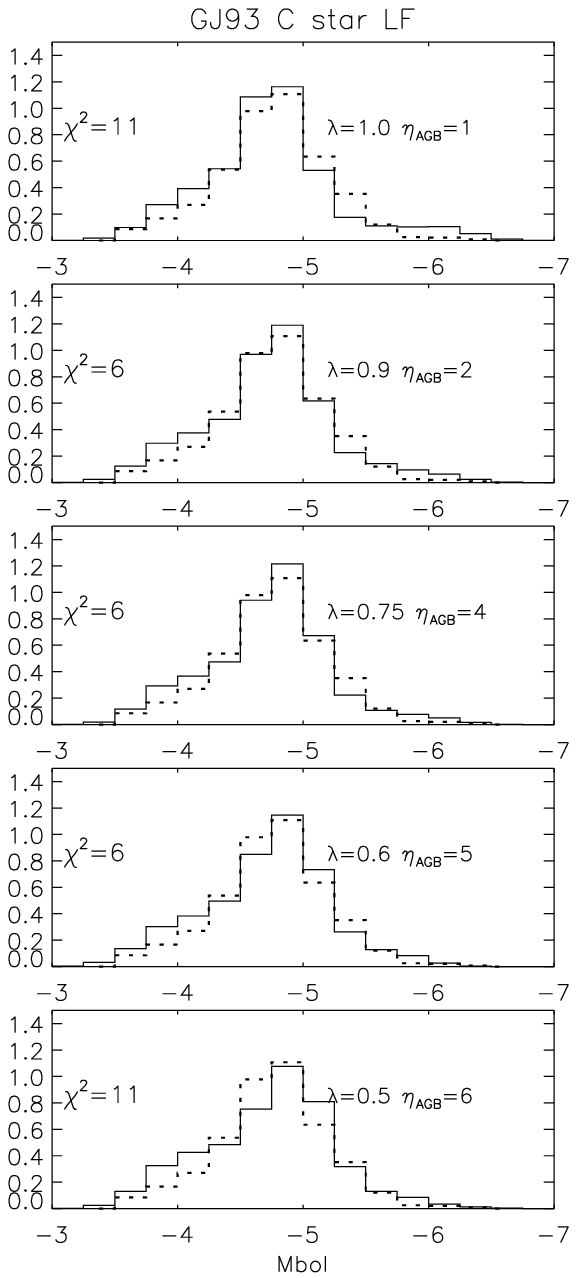,width=5.cm}}\quad
\subfigure{\epsfig{figure=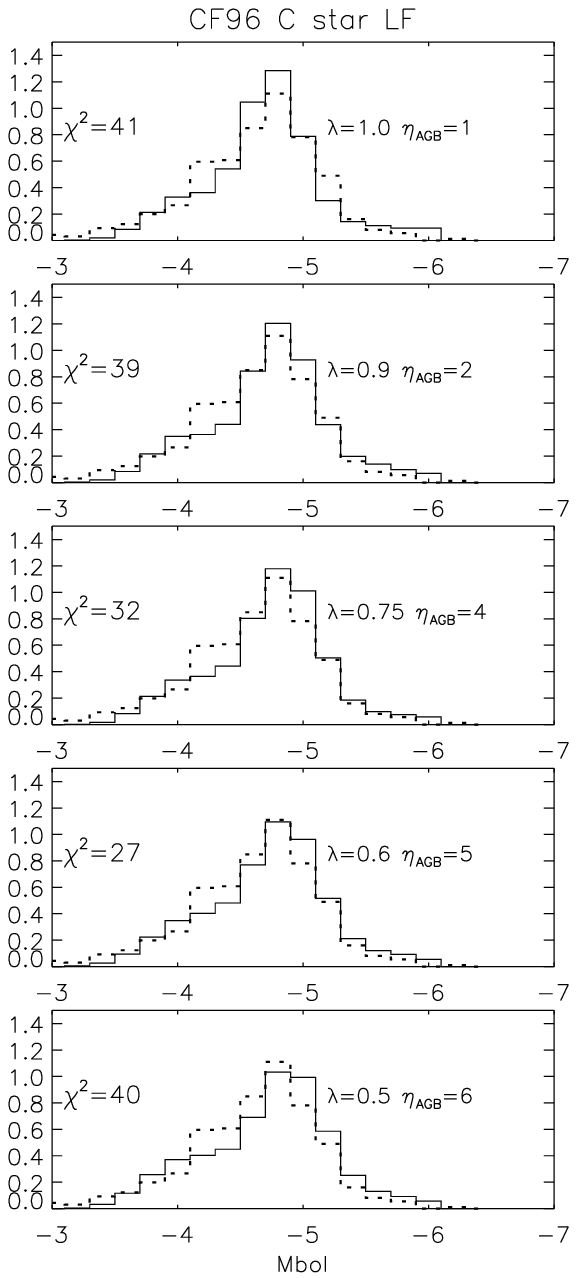,width=5.cm}}}
\caption{{\bf a}) The fraction of O-rich PNe in the solar
neighbourhood, in solid lines are model results from PN population
synthesis as a function of minimum progenitor mass (in
M$_{\odot}$). The coloured area corresponds to the observed fraction
$\pm1\sigma$.  $\eta_\mathrm{AGB}$ is the mass loss coefficient of a
scalable Reimers law, and $\lambda$ is the 3$^\mathrm{rd}$ dredge-up
parameter. {\bf b}) Synthesis (solid line) of the LMC C star LF used
in GJ93 (dotted line) , for the same sets of parameters as in
a).  {\bf c}) Synthesis (solid line) of the observed (dotted
line)  LMC C star LF  from Costa \& Frogel (1996).}\label{fig:model_sensitivity}
\end{figure*}

The predicted fraction of O-rich PNe is a sensitive function of the
free parameters, as shown in Fig.~\ref{fig:model_sensitivity}a. The
sets of parameters correspond to a range of 3$^\mathrm{rd}$ dredge-up
efficiencies, over an AGB star's lifetime (an extended lifetime
results in more dredge-up episodes).  But, at first sight of
Figs.~\ref{fig:model_sensitivity}b and \ref{fig:model_sensitivity}c,
all models seem to acceptably reproduce the C star LF used in GJ93 or
that from Costa \& Frogel (1996, CF96, with $1\sigma$ uncertainties of
$\sim 10\%$ at the peak). In terms of the C star LF $\chi^2$ values,
which we calculated summing over each logarithmic luminosity bin, from
4 to 5.8 for the CF96 LF and from 4.125 to 5.625 for the GJ93 LF, the
AGB parameters adopted by GJ93 ($\eta_\mathrm{AGB}=4$ and
$\lambda=0.75$) seem reasonable.  We assumed the GJ93 LF was built
with 298 sources, about three times less than in CF96 (895
sources). But the associated $\chi^2$ confidence levels are very low
(with 7 and 10 degrees of freedom for GJ93 and CF96, respectively).
It is difficult to assess the quality of the C star LF fit, from one
set of parameters to another. It turns out the C star LF $\chi^2$
values are very noisy: for the set of AGB parameters $(\lambda,
\eta_\mathrm{AGB}) = \left\{(1,1),\right.$  (0.9,2), (0.9,3), (0.6,3),
(0.75,4), (0.6,4), (0.75,5), (0.6,5),$ \left. (0.5,6) \right\}$, $\chi^2
=$(11, 6, 14, 7, 6, 8, 10, 6, 11) for the GJ93 LF and $\chi^2$
=(41, 39, 24, 56, 32, 39, 28, 27, 40) for the CF96 LF.  On the other hand,
f(O) exhibits broad variations over the same set of AGB parameters,
and the last two models with $\eta_\mathrm{AGB}=5$ are clearly
rejected. Models with $\eta_\mathrm{AGB}=2$ and $\lambda<0.6$ predict
excessive final masses (see Section~\ref{sec:mimf}), and
$M^\mathrm{min} \ls 1$ is unlikely considering the extremely low
proportion of type~I PNe (around 3\%). We stress that the constraints
put on the AGB parameters are model-dependent, rejected parameters
could well fit another set of AGB prescriptions. Our purpose here is
to illustrate the usefulness of PN population synthesis, provided a
proper PN sample is built.


Of the relations used in the synthetic AGB model (see Appendix), the
following include an explicit metallicity dependence: 
\begin{itemize}
\item the core-mass-interpulse period relation,
\item mass loss prior to the AGB, 
\item core-mass at the first thermal pulse,
\item luminosity at the first thermal pulse,
\item core-mass-luminosity relation,
\item $T_\mathrm{eff}(L,Z,M)$,
\item evolution rate,
\item hot bottom burning density and temperature.  
\end{itemize}
The synthetic PN population was tested for sensitivity on the galactic
metallicity gradient by fixing the metallicity entry of all the
previously listed relations at $Z=0.02$. The initial composition was
kept as described in the Appendix, i.e. a metallicity-scaled solar
mix, with a galactic metallicity gradient and an age-metallicity
relation.  The resulting synthetic PN population did not show a
gradient in the fraction of O-rich PNe. Subsequently the constraint on
the metallicity was relaxed one relation at a time, and the resulting
model population was inspected using the equivalent of
Fig. \ref{fig:results2}a. The difference in the fraction of O-rich PNe
inside and outside the solar circle was used to test the dependence
upon the metallicity gradient.


The most sensitive relations are $T_\mathrm{eff}(L,Z,M)$ and the
core-mass at the first thermal pulse.  The other relations resulted in
an insignificant change in the fraction of O-rich PNe inside and
outside R$_{\circ}$. The synthetic population generated using a flat
metallicity profile was confirmed not to show a gradient in the
fraction of O-rich PNe. The metallicity dependence of f(O) resides in
the following:
\begin{enumerate}
\item $T_\mathrm{eff}$ increases with lower metallicity, which for a
fixed luminosity translates into a lower envelope radius, and a lower
mass-loss rate. The extended AGB-lifetime thus leaves more time for
core-growth and the C-enrichment of the envelope. The relation used
here for $T_\mathrm{eff}$, is a model result from Wood (1990, see also
Vassiliadis \& Wood 1993).
\item The core-mass at the first thermal pulse, $M_c(1)$, is a
decreasing function of metallicity.  Together with a fixed minimum
core-mass for the occurrence of the third dredge-up (the free parameter
$M_c^\mathrm{min}$), the lower limit progenitor ZAMS mass for the
production of C-rich nebulae decreases with metallicity.
\end{enumerate}

A full search in all possible combinations in parameter space is
postponed to a future investigation, based on a more solid AGB model,
and a larger PN sample.

\section{Conclusions}\label{sec:conc_4}

PN population synthesis was investigated using a schematic model for
AGB evolution, whose free-parameters were calibrated with the
luminosity function of C stars in the LMC and the initial-final mass
relation. The galactic distribution of PN progenitors was generated
using the observed metallicity gradient and distribution of star
forming regions with galactocentric radius. The PNe with warm dust
emission from paper~I represent a homogeneous population, which is
presumably young and thus minimally affected by a possible dependence
of PN lifetime on progenitor mass. We examined the consequences of
assuming the statistics of PN compositions in the sample of paper~I
reflect tip-of-the-AGB values, and reached the following conclusions:
\begin{enumerate}
\item 
The fraction of PNe with O-rich dust places constraints on the range
of PN progenitor masses.  A minimum PN progenitor mass of
1\,M$_{\odot}$ predicts that about 50\% of all young PNe should be
O-rich, when only 22\% is observed. The minimum PN progenitor mass
required to reproduce the properties of the observed distribution of
PNe with warm dust is $M^\mathrm{min}=1.2$, at a $2\sigma$ confidence
level.
\item
The observed decrease in the fraction of O-rich PNe with galactocentric
radius is a robust property of the synthetic distribution. 
\item 
Independently of $M^\mathrm{min}$, values for the dredge up parameter
$\lambda$ and the mass-loss rate parameter $\eta_\mathrm{AGB}$ that
reproduce the LMC C star LF, do not necessarily match the observed
fraction of O-rich PNe, f(O). 
\end{enumerate}
The relative frequencies of C- and O-rich PNe depend mostly on the
value of the minimum core mass for C dredge-up, which is calibrated
with the LMC C star LF (as in GJ93). But the treatment of HBB is of
crucial importance in predicting PN compositions, as it directly
accounts for the Peimbert (1978) types. Measurements of the $^{12}$C
to $^{13}$C isotopic ratio in PNe can provide an observational test,
since a transition from $>$100 to $\sim$3 is expected with the onset
of HBB (e.g. RV81).

Overall, the good agreement obtained with the observed PN distribution
is an indication that the statistics of dust signatures in PNe can be
used as probes of their progenitor population in various galactic
environments.

\section*{Acknowledgments}

We are grateful to Mike Barlow, Robin Clegg, Edgardo Costa and Philipp
Podsiadlowski for interesting feedback and encouraging
discussions. S.C. acknowledges support from Fundaci\'{o}n Andes and
PPARC through a Gemini studentship.

\appendix 


\section{Analytical prescriptions for AGB evolution and calibration of
free parameters}

\subsection{Synthetic AGB evolution code}
\label{sec:app_sagb}

To summarise, we will describe the steps followed in the code, for a
star with fixed $(M_i,Z_i)$. All the analytical prescriptions are
written in GJ93, except when explicitly indicated as not being so.
The initial composition is calculated assuming the relative abundances
of the elements are the same as in the solar mix. We follow the steps
in GJ93 up to the first thermal pulse, taking the mass-loss
prescription on the E-AGB from Groenewegen et al. (1995, their Eq. 1).
The core mass and luminosity at the first thermal pulse, $M_c(1)$ and
$L(1)$, are the initial conditions for the TP-AGB.  $M_c(1)$ is of
significant importance, since it fixes for the most part the
initial-final mass relation, one of the observational
constraints. GJ93 take $M_c(1)$ from Lattanzio (1989) for $M_i$ under
$3~$M$_\odot$, and from RV81 above.  The subsequent TP-AGB evolution is
described by a loop, in which the interpulse period $t_\mathrm{ip}$ is
determined by a core-mass-interpulse relation (from Boothroyd \&
Sackmann 1988, GJ93 include a `turn-on' phase). The luminosity is
given by a core-mass-luminosity relation, with the fits in GJ93
(Boothroyd \& Sackmann 1988 and Iben \& Truran 1978), who use a
rectangular profile to account for the luminosity dip and flash.  GJ93
calculate the rate of core-growth from the luminosity due to hydrogen
burning\footnote{We did not include a luminosity increment due to HBB.
The observational constraint is the C star LF, and a HBB correction
would only affect massive M giants (and their mass loss rate), which
go in the C-phase at the very end of the AGB, see
Fig.~\ref{fig:tAGB}}.  The mass loss rate is estimated via a
{${\mathbf{\eta_\mathrm{AGB}}}$} scalable Reimers (1975) law.  The
stellar radius $R$ is estimated from the (model) relation given by
Wood (1990) between the luminosity and the effective temperature of
O-rich Miras. The condition for ending the AGB is then tested (based
on a critical envelope mass, Iben 1985), as well as the end of the
interpulse period. In the case of a new thermal pulse, the condition
for dredge-up is tested: $M_c>M_c^\mathrm{min}$.  $M_c^\mathrm{min}$
is one of the free parameters in GJ93.

The process following a third dredge-up event is the most sensitive
part of this code, and it affects directly the PN compositions. It is
standard to model the dredged-up material enriched in He and $^{12}$C
as a fraction ${\mathbf \lambda\sim{0.6}}$ of the core growth during
the interpulse period $\Delta{M_c}$,
\begin{equation}
\Delta{M}_\mathrm{dredged}= \lambda \, \Delta{M}_c.
\end{equation}
$\lambda$ could well be a function of metallicity or other parameters,
but is here taken as a constant. The composition of the dredged-up
material, the result of incomplete helium burning, is taken from
Boothroyd \& Sackmann (1988), $\mathrm{Y}^\mathrm{3rd-du}=0.76$,
$^{12}\mathrm{C}^\mathrm{3rd-du}=0.22$,
$^{16}\mathrm{O}^\mathrm{3rd-du}=0.02$. In the absence of HBB, the
dredged-up material is simply diluted in the envelope.

HBB is the main source of uncertainties in the model: existing AGB
models for single stars have difficulties in producing N and C rich
PNe (see Section \ref{sec:pb_N}).  HBB was treated by setting the
initial temperature $T_\mathrm{B}(1)$ and density $\rho_\mathrm{B}(1)$
at the bottom of the convective envelope to the values listed in
FC97. We first interpolated in $\log T_\mathrm{B}(1)$ and $\log
\rho_\mathrm{B}(1)$ as functions of $M_c(1)$, for the two metallicity
cases listed (LMC and solar), and then in $\log T_\mathrm{B}(1)$ and
$\log \rho_\mathrm{B}(1)$ as functions of metallicity $Z$. For the
subsequent evolution on the TP-AGB, we used the model results in FC97
as a function of time to estimate $dT_\mathrm{B}/dM_c$ and
$d\rho_\mathrm{B}/dM_c$.  They obtained constant growth rates for
$\rho_\mathrm{B}$ and $T_\mathrm{B}$, and we assumed core-growth to be
roughly linear in time (as it is in GJ93), and given by Table~5 in
FC97. A more accurate treatment is pointless, as major adjustments to
$dT_\mathrm{B}/dM_c$ are required to explain the observations. Thus
the results in FC97 led us to the following prescriptions for
$d\rho_\mathrm{B}/dM_c$ and $dT_\mathrm{B}/dM_c$, with upper limits
$d{\rho}_\mathrm{B}/dM_c < 10^{3}$~g cm$^{-3}$ M$_{\odot}^{-1}$,
$T_\mathrm{B} < 10^{8}K$, ${\rho}_\mathrm{B}<1$:
\begin{eqnarray}
\log \frac{dT_\mathrm{B}}{dM_c} & = & \left\{  \begin{array}{ll} 
8.33 M_c(1) + 2.05, &  Z=0.02, \\
6.85 M_c(1) + 3.51, &  Z=0.008, \\  \end{array} \right. \label{eq:dTbdM} \\  
\log \frac{d{\rho}_\mathrm{B}}{dM_c} & = & \left\{  \begin{array}{ll} 
14.3  M_c(1) -10.4,  &  Z=0.02, \\
13.1 M_c(1) - 9.23,  &  Z=0.008. \\  \end{array} \right.  \label{eq:drhobdM}
\end{eqnarray}

As explained in Section \ref{sec:pb_N}, a significant increase in
$T_\mathrm{B}$ during the AGB is required to explain C and N rich
PNe. Equation \ref{eq:dTbdM} does not provide such a strong increase,
except for the highest masses. As many uncertainties affect the
treatment of HBB anyway, we investigated treating $dT_\mathrm{B}/dM_c$
as a free parameter, and satisfactory results are obtained for
$dT_\mathrm{B}/dM_c= 2\,10^{9}$\,K\,M$_{\odot}^{-1}$.

Once $T_\mathrm{B}$ and $\rho_\mathrm{B}$ are determined, dredged up
material is exposed to CNO processing over a time $t_\mathrm{HBB}$,
and material in the envelope is processed during
$t_\mathrm{HBB}\,t_{ip} / \tau_\mathrm{conv}$, where
$\tau_\mathrm{conv}\sim$0.5\,yr is the convective turn-over
time-scale\footnote{The value of $\tau_\mathrm{conv}$ is of secondary
importance. The total exposure time is $t_\mathrm{HBB}\,t_{ip} /
\tau_\mathrm{conv}$, and any change in $\tau_\mathrm{conv}$ can be
compensated with $t_\mathrm{HBB}$.} (as quoted in FC97). We fixed
$t_{\mathrm{HBB}}=10^{-6}$yrs so as to obtain $^{12}$C/$^{13}$C ratios
at CN equilibrium values for the most massive stars, while avoiding
the onset of the full tri-cycle (which is not reported to occur in the
envelope of AGB stars, and reduces the O abundances to produce C/O
ratios of order 10). In the HBB treatment of GJ93, the envelope cannot
reach $^{12}$C/$^{13}$C at equilibrium values, since they took only a
very small fraction ($\sim10^{-4}$) of the envelope mass to be
CNO-processed. In GJ93 dredge-up material is fully exposed, with
$^{12}$C/$^{13}$C$\sim$3, but it is C-deficient, and $^{13}$C is
diluted in the envelope. It is worth noting that even with these
differences for HBB, the equivalent of Fig.~\ref{fig:mz} for the model
in GJ93 HBB is identical - except for $^{12}$C/$^{13}$C. The injection
of triple-alpha enriched material in the envelope is given by
\begin{eqnarray}
\Delta{Y^\mathrm{du}} & = & Y^\mathrm{3rd-du}\,\Delta{M}_\mathrm{dredged}, \\
\Delta{X_i^\mathrm{du}} & = & \frac{1}{t_\mathrm{HBB}}\int_0^{t_\mathrm{HBB}} X_i(t) dt \,\Delta{M}_\mathrm{dredged}, 
\end{eqnarray}
with initial conditions for the abundance of element $i$, $X_i(0)$
given as above, or otherwise taken as zero.  The change in the surface
abundances is 
\begin{equation}
X_i^\mathrm{new}= \frac{ \Delta{X_i^\mathrm{du}} +
\frac{M_\mathrm{env}}{t^\mathrm{env}_\mathrm{HBB}}
\int_{0}^{t^\mathrm{env}_\mathrm{HBB}} X_i(t) dt } { 
M_\mathrm{env}+\Delta{M}_\mathrm{dredged}} ,
\end{equation}
where $t^\mathrm{env}_\mathrm{HBB}=t_\mathrm{HBB}\, t_\mathrm{ip}/
\tau_\mathrm{conv}$, and the initial condition for the time average of
CNO processing is taken as $X_i(0)=X^\mathrm{old}$. The nuclear reaction
rates for the CNO tri-cycle were taken from Caughlan and Fowler
(1988), and we followed the analytical treatment in Clayton (1968),
with approximations for the branching ratios between the main CN-cycle
and the two secondary cycles from Podsiadlowski (1989).

As explained in Frost et al. (1998), there is a minimum envelope mass
below which HBB is quenched.  The critical envelope mass
$M\mathrm{^{env}_{HBB}}$ was adapted from GJ93,
\begin{eqnarray}
M\mathrm{^{env}_{HBB}}& = & 0.85 M_\mathrm{pn},\\
M_\mathrm{pn} & = & b\,(1.687-9.092\,M_c+11.687\,M_c^2\\ \nonumber
              &   & -4.343\,M_c^3),
\end{eqnarray}
where $M_\mathrm{pn}$ is the RV81 criterion for PN ejection with
b=0.5. The exact value of b and the 0.85 factor are of secondary
importance, GJ93 having fixed these parameters to match the results in
RV81. 

We stress here that dredge-up must occur after HBB is
quenched. Observational support can be found in Trams et al. (1999)
and van Loon et al. (1999).  It is of crucial importance for
explaining the vast majority of N rich PNe with C-rich dust
(corresponding approximately to the last 2000~yr of AGB mass loss).


\subsection{Observational constraints}\label{sec:obs_cal}
\label{sec:app_obs}

\subsubsection{The luminosity function of C stars in the LMC}
\label{sec:lmctest}

The population of stars in the LMC with initial masses between $M$ and
$M+dM$ which are currently on the TP-AGB is
\begin{equation}
N(M)dM \propto \mathrm{SFR}(M) \times \mathrm{IMF}(M) \times t_\mathrm{AGB}(M)dM ,
\end{equation}
where $\mathrm{SFR}(M)$ is the star formation rate in the LMC at time
$t(M)$, the time of birth as a function of mass, $\mathrm{IMF}(M)$ is
the initial mass function, which will be taken as a power-law, and
$t_\mathrm{AGB}(M)$ is the lifetime on the TP-AGB as a function of
main-sequence mass. There is an implicit metallicity dependence via
the LMC age-metallicity relation.  In practice, we started with
$N(M)dM \propto \mathrm{SFR}(M) \times \mathrm{IMF}(M)$, and weighted
the result with $t_\mathrm{AGB}(M)$. The procedure is equivalent to
multiplying each mass entry by $t_\mathrm{AGB}(M)$ (times an integer
big enough to keep a satisfactory resolution).

The parameters describing the global evolution of the LMC were
reproduced from GJ93.  The star formation rate in the LMC is
approximated as $\mathrm{SFR}(M) \propto \exp(-t(M)/7)$, with $t(M)$
in Gyr and an LMC age of $\tau_\mathrm{LMC}$=11\,Gyr, the initial mass
function is $\mathrm{IMF}(M) \propto M^{-2.72}$, and the
age-metallicity relation is assumed to be linear in time, $Z(t) =
0.01-0.008\times(1-t/13)$.  The total lifetime as a function of mass,
$t(M)$, was taken from Iben \& Laughlin (1989).

\begin{figure}
\begin{center}
\resizebox{8cm}{!}{\epsfig{file=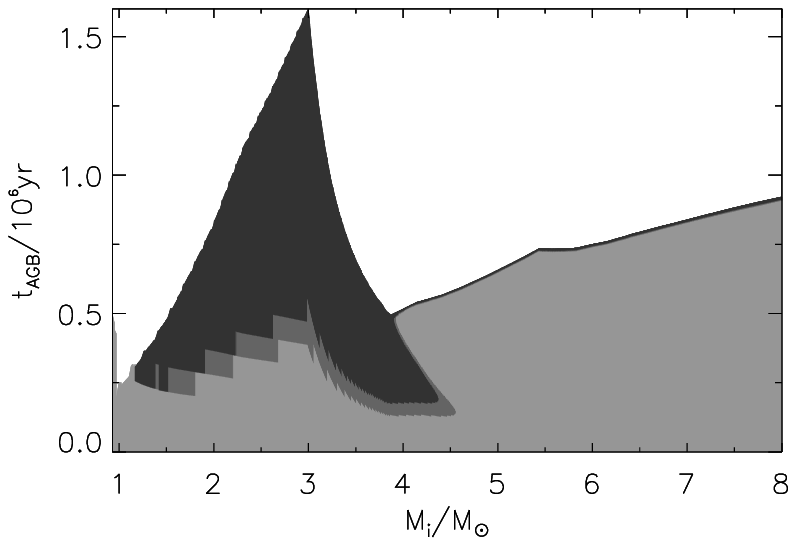}}
\end{center}
\caption{ Time spent on the AGB (10$^{6}$ years, in ordinates) as a
function of initial mass, for the LMC model. The grey scale grows
darker from the M to S to C phases.}\label{fig:tAGB}
\end{figure}

Observational data with which to constrain the AGB population in the
LMC is compiled in GJ93. The luminosity function of C stars is a very
good constraint since there can be no doubt on their evolutionary
status (i.e., at the luminosity implied by the LMC distance, C stars
can only be on the TP-AGB). On the other hand the relative frequency
of C and M stars is much more uncertain: the contamination from M
stars on the E-AGB is difficult to estimate. Hughes (1989) and Hughes
\& Wood (1990) found that most LPVs in the LMC have spectral type M5+,
and C/M$=0.63$, provided all M stars on the TP-AGB are also
LPVs. Based on previous measurements (Blanco \& McCarthy 1983), GJ93
favour C/M$>0.63$.  But M giants spectral types are difficult to
assess, and we feel it is safer to trust the C star counts, and admit
that the number of M stars on the TP-AGB is not known, although
probably greater than the number of O-rich LPVs - thus C/M$\ls
0.6$. With $t_\mathrm{M}$, $t_\mathrm{S}$, $t_\mathrm{C}$, the time
spent by a star in the M, S and C phases, the C/M giants ratio on the
TP-AGB is given by
\begin{equation}  \label{eq:CM}
\mathrm{C/M}= \frac{\int dM \, \mathrm{SFR}(M) \times \mathrm{IMF}(M) \times
t_\mathrm{C}(M)}{\int dM  \, \mathrm{SFR}(M) \times \mathrm{IMF}(M) \times
t_\mathrm{M}(M)},
\end{equation}
where the integral runs from $0.93\,$M$_{\odot}$ to $\sim 8\,$M$_{\odot}$
(we kept the lower mass limit in GJ93).  The M phase corresponds to
C/O$<$0.81, the S phase to 0.81$<$C/O$<$1, and the C phase to C/O$>$1
(Smith \& Lambert 1986). We followed GJ93 in neglecting obscuration of
bright C stars due to excessive mass loss. GJ93 also estimated that
not more than 3\% of C stars brighter than
$\mathrm{M}_\mathrm{bol}=-6$ could have been missed by the optical
surveys. The functions $t_\mathrm{M}$, $t_\mathrm{S}$, $t_\mathrm{C}$
are shown in Fig. \ref{fig:tAGB}.  The formula above, as well as the
Monte Carlo runs, predict that M, S and C stars should occur in the
ratio .65:.04:.31. This is a (TP-AGB) C/M star ratio of 0.47, in rough
agreement with observations - provided the observed ratio is indeed
about $0.6$; the LMC C/M star ratio on the TP-AGB is a rather
uncertain quantity. 


The observed LF of C stars in the LMC was reproduced from Costa \&
Frogel (1996). Fig.~\ref{fig:model_sensitivity} gives examples of
synthetic LFs that approximate the observed LF.  The synthetic LFs
were smoothed with a $\sigma=0.2$ gaussian, as in GJ93. Structure on
small scales would not be picked up in observations anyway; there is
considerable scatter about the bolometric correction law (Costa \&
Frogel 1996, their Fig.~1).  The distance modulus to the LMC was taken
as 18.5.


As a note of caution, Marigo et al. (1999) have shown how the C star
LF is more sensitive on metallicity than it is on either the star
formation history or the IMF.  The metallicy dependence of the
prescriptions we use also results in such a sensitivity, but an
extrapolation from the LMC to the solar neighbourhood may seem
far-fetched. It is assumed that none of the free parameters are
metallicity dependent.

\subsubsection{The initial-final mass relation} \label{sec:mimf}

\begin{figure}
\begin{center}
\resizebox{8cm}{!}{\epsfig{file=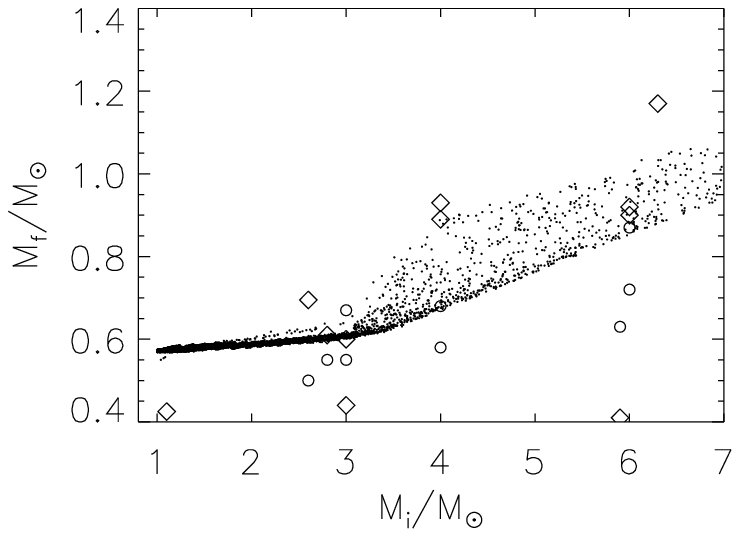}}
\end{center}
\caption{The initial-final mass relation from the galactic model in
this article, and as observed.  Diamonds and open-circles correspond
to final mass estimates based on surface-gravity or stellar radius,
respectively.}\label{fig:Mi_Mf}
\end{figure}

The initial-final mass for the `standard' galactic model is shown in
Fig.  \ref{fig:Mi_Mf}, where it is compared to the data from Weidemann
\& Koster 1983 and Weidemann 1987 (data points from FC97).  Final
masses are derived from the surface gravity, or the stellar radius;
there are thus sometimes two data points per object, with the same
$M_i$. The agreement is satisfactory, except at low initial masses,
where the model initial-final mass relation might be an overestimate
(it should be kept in mind that the `observed' initial-final mass
relation is in fact model dependent). Wagenhuber \& Groenewegen (1999)
reach the conclusion that the GJ93 approximations for the core masses
at the first thermal pulse (which are used here, and largely determine
the final mass), overestimated most other model results at low initial
mass, but the discrepancy is small. The scatter in the model $M_i-M_f$
stems from the metallicity dependence of the analytical prescription
for the core mass at the first thermal pulse (it is smaller for the
prescription in Wagenhuber \& Groenewegen 1999).

\subsection{Evolution on the TP-AGB for characteristic initial masses}

\begin{figure}
\centering
\mbox{\subfigure{\epsfig{figure=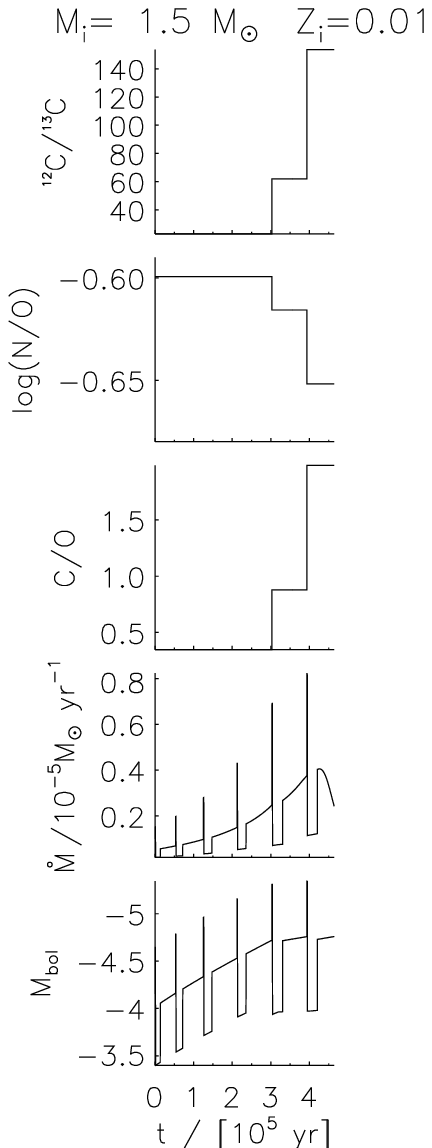,width=2.63975cm}}\quad
\subfigure{\epsfig{figure=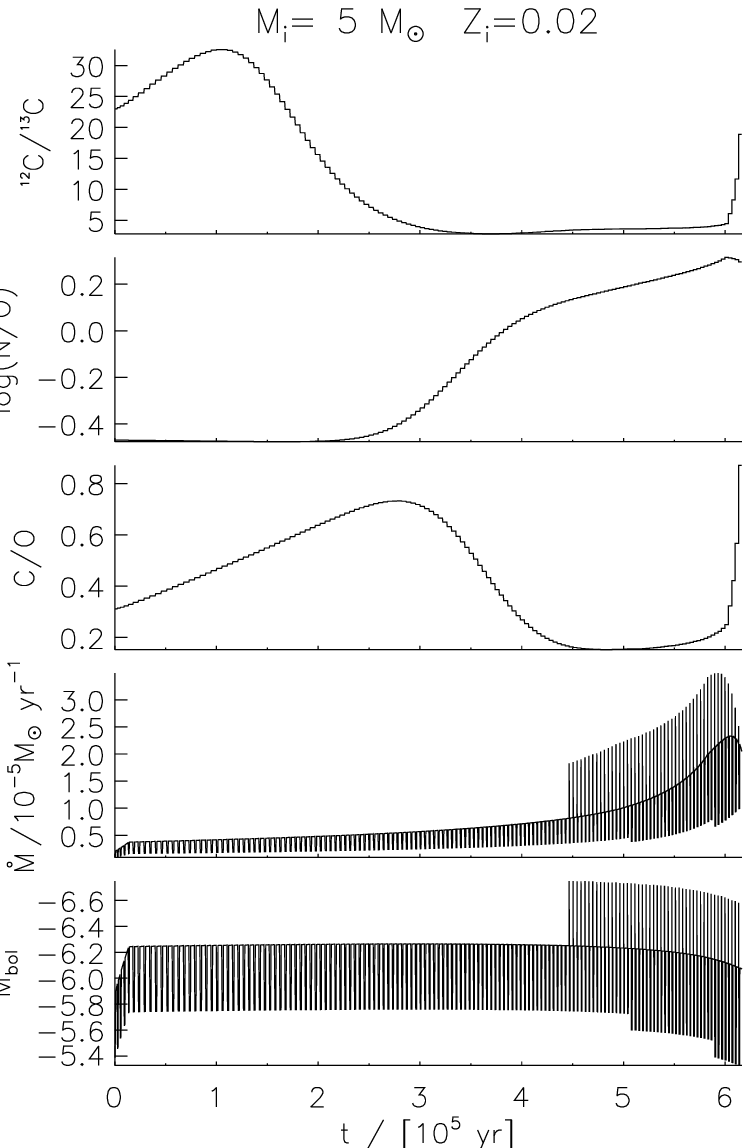,width=5.86cm}}}
\caption{ Synthetic evolution for characteristic initial masses. The
results for the surface $^{12}$C/$^{13}$C, log(N/O), C/O, the mass
loss rate and bolometric absolute magnitude are shown from top to
bottom, for initial masses 1.5 and 5~M$_{\odot}$ from left to
right).}\label{fig:ex_sagb}
\end{figure}

The synthetic AGB evolution of models with $(M_i,Z_i)=\left\{
(1.5,0.01);(5,0.02)\right\}$ are shown in figure
\ref{fig:ex_sagb}. The case with $M_{i}=1.5\,$M$_{\odot}$ shows how
towards the low mass end of C-rich PNe progenitors, a low number of
thermal pulses gradually builds up a $^{12}$C surface
overabundance. With $M_{i}=5\,$M$_{\odot}$, the first $\sim
2~10^{5}$~yr correspond to surface C enrichment, until core growth is
high enough for $T_\mathrm{B}$ to reach CNO processing values.  The
total number of thermal pulses reaches 143, and at
$t=10^{6}\,\mathrm{yr}$ the envelope mass drops below the adopted
critical mass for the quenching of HBB.


\bsp
\label{lastpage}

\begin{thebibliography}{}
\bibitem{} Aller L.H., Czyzak S.J., 1983, ApJSS, 51, 211
\bibitem{} Barlow M.J., 1983, IAU symp 103 ``Planetary Nebulae'' ed D.R. Flower p105
\bibitem{} Blanco V.M., McCarthy M.F., 1983, AJ, 88, 1442
\bibitem{} Boothroyd A.I., Sackman I.J., 1988, ApJ, 328, 653
\bibitem{} Bronfman L., Casassus S., May J., Nyman L.A., 2000, A\&A,
358, 521
\bibitem{} Bronfman L., Cohen R.S., Alvarez H., May J., Thaddeus P., 1988, ApJ, 324, 248
\bibitem{} Bryan G.L., Volk K., Kwok S., 1990, ApJ, 365, 301
\bibitem{} Carraro G., Yuen Keong N.,  Portinari L., 1997, astro-ph/9707185
\bibitem{} Casassus S., Roche P.F., Aitken D.K., Smith C.H.,  2000,
MNRAS, {\em in press}
\bibitem{} Caughlan G.R., Fowler W.A., 1988, Atomic Data and Nuclear Data Tables, 40, 2
\bibitem{} Clayton D.D., 1968, `Principles of Stellar Evolution and Nucleosynthesis', New York: McGraw-Hill
\bibitem{} Condon J.J., Kaplan D.L., 1998, ApJSS, 117, 361
\bibitem{} Costa E., Frogel J.A., 1996, AJ, 112, 2607
\bibitem{} Deul E.R., Burton W.B., 1991, in {\em The Galactic Interstellar Medium}, Saas Fee Advanced Course 21, p 79
\bibitem{} Fa\'{u}ndez-Abans M., Maciel W.J., 1987, A\&A, 183, 324
\bibitem{} Forestini M, Charbonnel C., 1997, A\&AS, 123, 241 (FC97)
\bibitem{} Frost C.A., Cannon R.C, Lattanzio J.C., Wood P.R., Forestini M., 1998, A\&A, 332, L17
\bibitem{} Groenewegen M.A.T, deJong T., 1993, A\&A,  267, 410 (GJ93)
\bibitem{} Groenewegen M.A.T, van den Hoek L.B., de Jong T., 1995, A\&A,
293, 381 
\bibitem{} Henry R.B.C., Kwitter K.B., Bates J.A., 2000, ApJ, {\em in press} 
\bibitem{} Hughes S.M.G., 1989, AJ, 97, 1634
\bibitem{} Hughes S.M.G., Wood P.R., 1990, AJ, 99, 784
\bibitem{} Iben I. Jr., 1985, QJRAS, 26, 1 
\bibitem{} Iben I. Jr., Laughlin G., 1989, ApJ, 341, 312
\bibitem{} Iben I. Jr., Truran J.W., 1978, ApJ, 224, L63.
\bibitem{} Jura M., Joyce R.R., Kleinmann S.G., 1989, ApJ, 336, 924
\bibitem{} Kaler J.B., Iben I. Jr., Becker S.A., 1978, ApJ, 224, L63 
\bibitem{} K\"{o}ppen J., Cuisinier F., 1997, A\&A, 319, 98
\bibitem{} K\"{o}ppen J., Vergeley J.-L., 1998, MNRAS, 299, 567
\bibitem{} Lattanzio J.C., 1989, ApJ, 347, 989
\bibitem{} Leisy P., Dennefeld M., 1996, A\&ASS, 116, 95 
\bibitem{} Maciel W.J., Dutra C.M., 1992, A\&A 262, 271
\bibitem{} Marigo P., Bressan A., Chiosi C., 1996, A\&A, 313, 545
\bibitem{} Marigo P., Girardi L., Bressan A., 1999, A\&A, 344, 123
\bibitem{} Meusinger H., Reimann H.-G., Stecklum B., 1991, A\&A, 245, 57
\bibitem{} Peimbert M., 1978, in Terzian Y., ed., Proc. IAU Symp 76, `Planetary Nebulae'. Reidel, Dordrecht, p. 233
\bibitem{} Podsiadlowski P., PhD Thesis, 1989, M.I.T.
\bibitem{} Ratag M.A., Pottash S.R., Dennefeld M, Menzies J., 1997,  A\&ASS, 126, 297
\bibitem{} Reimers D., 1975, in `Problems in Stellar Atmospheres and Envelopes', Basheck B. et al. eds., Springer, Berlin, p 229
\bibitem{} Renzini A., Voli M., 1981, A\&A, 94, 175 (RV81)
724 
\bibitem{} Roche P.F., 1989, IAU symp 131 ``Planetary Nebulae'' ed S.Torres Peimbert, p117
\bibitem{} Sabas V., 1997, Proceedings of the ESA Symposium `Hipparcos-Venice '97', ESA SP-402, 563
\bibitem{} Siess L., Livio M., 1999, MNRAS, 3304, 925 
\bibitem{} Smith V.V., Lambert D.L., 1986, ApJ 311, 843
\bibitem{} Thronson H.A., Latter W.B, Black J.H., Bally J., Hacking P., 1987, ApJ, 322, 770
\bibitem{} Trams N.R., et al., 1999, A\&A, 344, L17
\bibitem{} van Loon J.Th., Zijlstra A.A., Groenewegen M.A.T., 1999,
A\&A 346, 805
\bibitem{} Vassiliadis E., Wood P.R., 1993, ApJ, 413, 641
\bibitem{} Wagenhuber J., Groenewegen M.A.T., 1999, A\&A, 340, 183
\bibitem{} Weidemann V., 1987, A\&A, 188, 74
\bibitem{} Weidemann V., Koester D., 1983, A\&A, 121, 77
\bibitem{} Wielen  R., 1977, A\&A, 60, 263
\bibitem{} Wielen  R., Fuchs B., Dettbarn C., 1996, A\&A, 60, 263
\bibitem{} Wood P.R., 1990, in Mennessier M.O., Omont A. (eds), `{\it From Miras to Planetary Nebulae}', Editions Fronti\`{e}res, Gif-sur-Yvette, p67 
\bibitem{} Zuckerman B., Aller L.H., 1986, ApJ, 301, 772 
\end{thebibliography}
\end{document}